*energies*

*Draft article submitted and accepted to Energies*

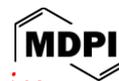

*Article*

# Empirical validation of a thermal model of a complex roof including phase change materials.

**Stéphane GUICHARD [1,†], Frédéric MIRANVILLE [2,†], Dimitri BIGOT [2,†], Bruno MALET-DAMOUR [2,†], Teddy LIBELLE [2,†] and Harry BOYER [2,†]**

[1]  Research Institute in Innovation and Business Sciences (IRISE) Laboratory \CESI-CCIR\CRITT, The CESI engineering school, Campus Pro – CCIR 65 rue du Père Lafosse-Boîte n°4 97410 Saint-Pierre, FRANCE.;

[2]  University of Reunion, Physics and Mathematical Engineering Laboratory for Energy, Environment and Building (PIMENT), 117, rue du Général Ailleret 97 430 Le Tampon, FRANCE.;
    frederic.miranville@univ-reunion.fr   (F.M.); dimitri.bigot@univ-reunion.fr (D.B.);
    bruno.malet-damour@univ-reunion.fr (B.M.-D.); teddy.libelle@univ-reunion.fr (T.L.);
    harry.boyer@univ-reunion.fr (H.B.);

*   Correspondence: sguichard@cesi.fr (S.G); Tel.: +262-0262-700-734

†   These authors contributed equally to this work.

**Abstract:** This paper deals with the empirical validation of a building thermal model using a phase change material (PCM) in a complex roof. A mathematical model dedicated to phase change materials based on the heat apparent capacity method was implemented in a multi-zone building simulation code, the aim being to increase understanding of the thermal behavior of the whole building with PCM technologies. To empirically validate the model, the methodology is based both on numerical and experimental studies. A parametric sensitivity analysis was performed and a set of parameters of the thermal model have been identified for optimization. The use of a generic optimization program called GenOpt® coupled to the building simulation code enabled to determine the set of adequate parameters. We first present the empirical validation methodology and main results of previous work. We then give an overview of GenOpt® and its coupling with the building simulation code. Finally, once the optimization results are obtained, comparisons of thermal model of PCM with measurements are found to be acceptable and are presented.

PACS: J0101

## 1. Introduction

Buildings are indisputably considered as one of the largest energy consuming sector. According to the International Energy Agency (IEA), the average energy consumed by buildings represents 32% of energy consumption worldwide, with approximately 40% of primary energy used in most countries. In France, the energetic consumption in the building sector is approximately about 43%, representing a quarter of national's carbon dioxide emissions. The use of energy-hungry appliances to improve the thermal comfort is responsible both for the electrical energy consumption and the increase of greenhouse gas emissions [1].

For this reason, some actions are led to curb energy consumption and to protect environment, for example, the use of renewable energies, passive energy buildings and the use of building codes for new or retrofited buildings. Moreover, several studies and applications have shown that the building thermal inertia is among possible solutions and should be improved, in order to reach high performance and low-energy buildings [2–4]. As a result, the question in the building sector arises as to the increase of the thermal energy capacity storage in the used materials.. With this target in mind, a technology such as the phase change materials may be integrated into building envelopes, both to enhance the thermal energy storage [5] and to improve the thermal comfort. Indeed, because of their higher thermal energy storage densities than other heat storage materials, these materials are able to store and release thermal energy as latent heat, when the phase change occurs. It is important to highlight the fact that the latent heat storage is much larger than the sensible heat storage [6]. Usually, organic and inorganic PCMs are often used and solid-liquid phases are chosen [7]. Despite its thermal conductivity that should be improved, paraffin is often used for latent heat thermal energy storage [8]. Indeed, it has useful thermal properties such as absence of super cooling, chemical stability and low vapor pressure [3,9].

The use of PCMs in the building envelope may reduce the peak loads and heating, ventilating and air conditioning (HVAC) energy consumptions by increasing the thermal inertia of each walls of the building. Indeed, the peak load may be shifted to the off peak load periods of energy use [10]. In addition, the results of PCM-oriented research on buildings have shown that the thermal comfort was improved and energy saving could be made.

Among all the PCM applications to reach high energy efficiency buildings, for example PCM integrated wall, PCM assisted ceiling heating and cooling, photovoltaic system coupled with a PCM-based heat storage [11], and so on, this papers deals with the inclusion of PCMs in the roof system. Generally, the roof is considered as a thermal buffer between the indoor and outdoor environment. This is the part of the building most exposed to solar radiation in hot climates and considered as the weakest part of the building thermal performance [12,13].

A possible solution to reduce heat transfer from outside to inside of the building may be found in the increased use of mass insulation. Nevertheless, this type of thermal insulation allows to reduce heat transfer due to conduction, but does not reduce heat transfer by infrared radiation through the roofing. To overcome, a solution is based on the use of radiant barrier systems, which requires the presence of air layers in order to benefit a reflective radiation [12]. Based on the same approach, PCM under a flexible sheet form laminated both sides with an aluminum sheet is used in the lower part of the roof between the air layer and the drywall, both to enhance the thermal energy storage and to reduce the solar radiation through the roof system [14].

Furthermore, before integrating PCM into new or retrofit buildings, it will be interested to predict the thermal effects of these materials on the whole building. To contribute to the energy efficiency policy recommendations on buildings energy consumption, a one-dimensional simplified numerical model for phase change material has been developed and implemented in a prototype of building simulation code named ISOLAB. This one is able to take into account the building envelopes, including PCM or not, and the actual impact on energy consumption. Nevertheless, before using PCM model integrated in ISOLAB code and to ensure the reliability of the results, this article focuses on the empirically validation of the latter. We first present the tools used, building simulation code and the generic optimization program GenOpt®. We then briefly describe the studied system as well as main results of previous work. Finally, results of the experimental validation of the thermal model are presented and discussed.

## 2. Presentation of tools

### 2.1. A building simulation code: ISOLAB

Developed by MIRANVILLE, ISOLAB is a prototype of building simulation software developed with the Matlab environment [15]. Nodal description of buildings and finite difference scheme of the time variable in one-dimensional are used to simulate the dynamic thermal behavior of a monozone or multi-zone building according to its environment (weather data and location). In order to determine temperatures of the whole building, the following matricial formalism (1) is solved by using an implicit finite-difference method [16].

$$C_w \dot{T}_w = A_w T_w + B_w, \tag{1}$$

Where the index w is used for walls and windows. $A_w$ is the state matrix including the terms linked to heat conduction and the interior linearized convective exchanges. $B_w$ is the vector containing outside or internal solicitations of the system. $C_w$ is the capacities matrix. $T_w$ is the state vector containing every temperatures of each wall, $\dot{T}_w$ and is the temperature derivative of $T_w$.

ISOLAB code has already been validated with the IEA BESTEST procedure (International Energy Agency, Building Energy Simulation Test), concerned with buildings without PCMs [15].

### 2.2. A generic optimization program: GenOpt®

The optimization process consists in running several model simulations with different parameters set. The chosen parameters are identified by performing a mathematical tool like parametric sensitivity analysis. Thereafter, each simulation is launched with a different set of parameters (continuous parameters, discrete parameters, or both), and the associated cost function value, with or without constraints, is recorded. Finally, to ensure convergence to the best set of parameters, it is necessary to run many simulations on each parameter for the right different parameter range.

A generic optimization program, called GenOpt® from the Lawrence Berkeley National Laboratory of University of California, developed with the Java environment by Mickael Wetter is used to optimize the unknown parameters [17]. The choice of GenOpt® has also been made according to its simulation program interface that allows the coupling with any simulation program, without requiring code modifications. Indeed, the coupling is done by creating some files required by GenOpt® to run an optimization and also an auto-executable version of the model [18]. In addition, GenOpt® is dedicated for the thermal building simulation. For more details about the generic optimization program, the interested reader may refer to [17].

Before performing the optimization sequences with GenOpt®, the studied system with the main results of previous work are presented in the following paragraph.

## 3. Experimental set up of a complex standard roof including phase change materials

### 3.1 Introduction

The objective of this part is to show the main results of experimental investigations on a complex roof incorporating PCM panels. The use of such a database allows to empirically validate the numerical code developed and implemented in a prototype of building simulation software to determinate the thermal performance of any building envelopes with PCM.

First, a description of the roofing complex and the different associated instrumentation is presented in the following paragraph.

### 3.2. Localization and structure of the test cell

With dimensions of 3m (height) × 3m (width) × 3m (length) and an internal volume about 30 m³, the test cell, called LGI, can be considered as a typical room of buildings existing in Reunion Island. It has been designed according to a flexible structure in order to study several configurations and phenomena. Thanks to its modular structure, the movable walls allow to include PCM panels to the standard roof inclined at 20° to the horizontal. The whole building envelopes are constituted of vertical opaque walls, a jalousie, a glass door and a complex roof with PCM panels. The roofing complex is an assembly of homogeneous or inhomogeneous materials, separated by one or several air layers [12,14,18]. The building components of LGI are given in Table 1, and an overall view is depicted in Figure 1a.

**Table 1.** Arrangement of LGI test cell [13]

| Element | Composition | Remark(s) |
|---|---|---|
| Opaque vertical walls | Sandwich board 80mm thick cement-fiber/polyurethane/cement-fiber | |
| Window | Aluminium frame, 8 mm clear glass | Blind-type 0.8x0.8m |
| Glass door | Aluminium frame, 8mm clear glass | Glass in upper and lower parts, 0.7x2.2m |
| Roofing complex | Corrugated galvanised steel 1 mm/air layer of 280 mm thick/PCM 5.26 mm thick/ Plasterboard 12.5 mm | PCM is laminated to aluminium protective foils. Roof inclined at 20° |
| Floor | Concrete slabs 80mm thick on 60 mm thick polystyrene | |

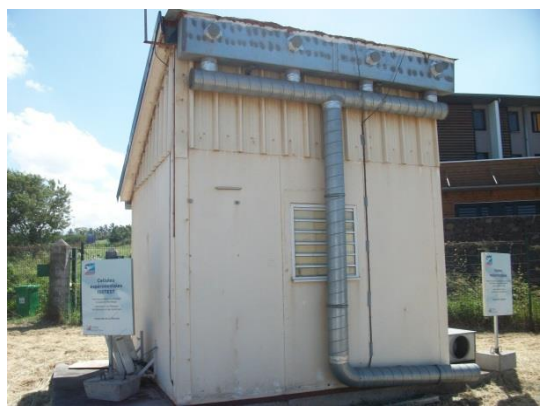
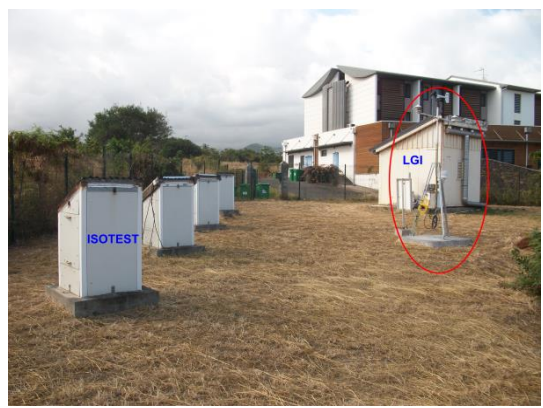

(a)                                      (b)

**Figure 1.** (**a**) LGI cell; (**b**) Experimental platform.

Located at a low altitude sea level (55 m), the experimental set up was erected on the experimental platform of the University Institute of Technology of Saint-Pierre (Reunion Island). The choice of this location results in a tropical climate with strong solar radiation and humidity. Besides, the test cell is oriented north in order to receive symmetrical solar radiation during the day [14]. With the aim of obtaining extreme solicitations input from the roof, a dark blue color has been chosen for the corrugated iron. During the procedure of experimental validation, the blind windows and the window panes were masked, and the test cell was kept closed, without using mechanical ventilation or air-conditioning system [13].

Furthermore, near to the experimental devices, two meteorological stations (a weather station in red circle is depicted in Figure 1b) are installed in order to collect data from the environment, such as solar radiation (global, direct and diffuse, on a horizontal plane), ambient air temperature, exterior relativity humidity, wind speed and direction. As depicted in Figure 2 and because of the important number of days (92 days approximately) of the chosen experimental sequence and, in order not to overload the graphs, only four days of data from August to October 2012 are presented in this paper.

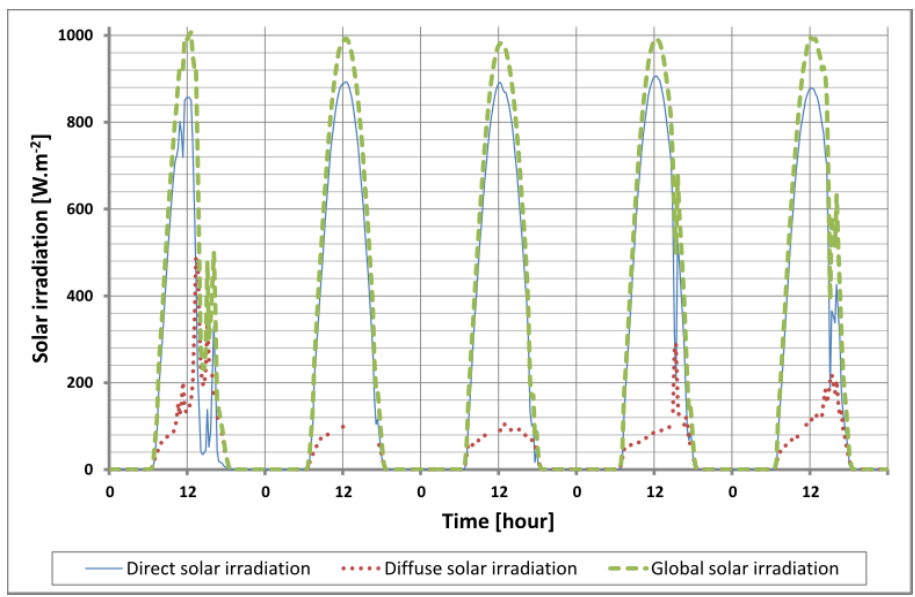

**Figure 2.** Solar irradiation from 20th September to 24th September 2012 [13].

### 3.2.1. Instrumentation of the enclosure

Located both in the enclosure and on the roof of the LGI test cell, sixty sensors have been used.

The walls (north, south, east and west, inside and outside) are equipped with thermal sensors on surfaces, such as T-type thermocouples and flux meters in order to measure the inside or outside surface temperature of each wall. For air temperature, the thermocouples are inserted into aluminum cylinders. For radiant temperature, the thermocouples are contained in a black globe.

To assess the stratification of the air, the interior volume was measured at three different heights from floor to ceiling (see Figure 3). Many heat sensors have been sealed in the concrete slabs supporting the cell, in order to assess the information on the boundary conditions from the ground. Indeed, the ground model is difficult to develop and to avoid modeling errors concerning this part of the buildings not directly linked to our study, the database from the ground measures will be used during the code validation step.

For more details on sensors location and set-up and the on the associated errors, the interested reader may refer to [13].

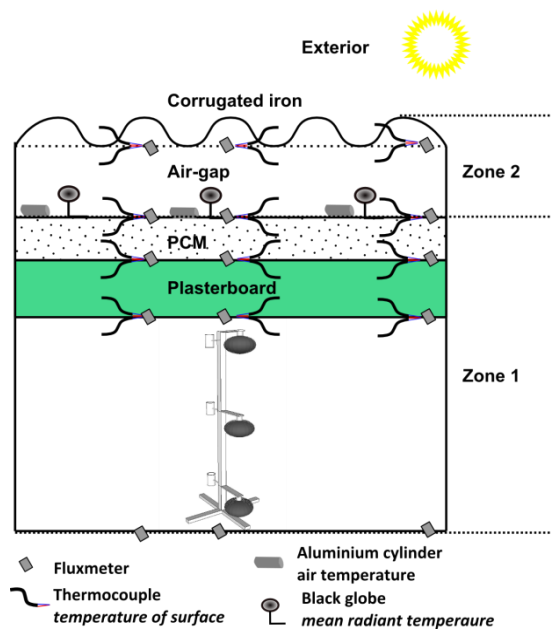

**Figure 3**. Simplified scheme of LGI cell and instrumentation

### 3.2.2. Instrumentation of the roof

To ensure reliable measurements from the complex roof components, sensors have been located as depicted in Figure 3. To assess the temperatures and the heat flux meters from to the corrugated iron to the ceiling, each component of the complex roof was instrumented. The heat sensors are spread over on both sides of the surfaces of the corrugated iron, on PCM panels and on drywall. Between PCM panels and the corrugated iron, the air layer is not ventilated, and the dry-air and black globe temperatures were also measured.

Before using heat sensors, the thermocouples were calibrated and verified on site according to protocol dictates. Heat flux sensors were calibrated by manufacturer. The accuracy of the thermocouples is estimated to ±0.5°C, and according to manufacturer, the relative error of flux meters is approximately about 5%. In order to avoid air bubbles for both thermocouples and flux meters, a conductive heat paste was applied on the surfaces.

A data logger was installed in the LGI cell in order to collect data from all sensors every 15 min. All data were saved on a computer [13].

### 4. Previous investigation

#### 4.1. Introduction

Since May 2013, both the thermal behavior of building envelopes including PCM and time-varying thermal properties of materials were not taken into account by ISOLAB. For this, a simplified numerical model of the thermal behavior of phase change materials was developed and implemented. However, many constraints on the PCM model have been imposed, such as the respect of the state system formalism and the use of implicit scheme in one-dimension according to the finite difference approach. The apparent heat method capacity has moreover been used.

#### 4.2. Description of the roofing complex

The model is intended to predict temperature behavior of each component of the whole building, including the inside air of enclosures. The studied LGI cell is considered as well-isolated and divided into two thermal zones. Indeed, the ceiling is a specific wall separating two zones. It plays the role of ceiling for the first zone and of floor for the second zone. We note that the air gap simulated in zone 2 is considered as a thermal zone in our multi-zone building model (see Figure 3).

The presence of an air layer between the corrugated iron and PCM panels allows to benefit the principle of action of reflective insulation, which is closely linked to the radiative properties of the surfaces of PCM panels. Taking into account the air layer with the combination of homogenous and inhomogeneous materials, the roof system can be qualified as complex with coupled heat transfers involved (conduction, convection and radiative transfers) [19]. Therefore, the chosen configuration may complicate the determination of thermal performances due to the multiple configurations of the air layer: opened or closed, naturally ventilated or forced-ventilated [14]. This is the raison explaining that in the proposed approach in this paper, the enclosed air space is considered as a thermal zone.

### 4.3. Description of PCM test

PCM tested is the commercial product from Dupont™ Energain®. It is a flexible sheet 5 mm thick, made of 60% microencapsulated paraffin wax within a copolymer laminated on both sides with an aluminum sheet [13,14]. Its characteristics are summarized in the following Table 2:

**Table 2**. Characteristics of PCM used [13]

| Parameter | Value | Unit |
|---|---|---|
| Thermal properties | | |
| Thermal conductivity: $\lambda_s/\lambda_l$ | 0.22/0.18 | W.m⁻¹.K⁻¹ |
| Heat capacity: $C_{ps}/C_{pl}$ | 3134/2833 | J.kg⁻¹.K⁻¹ |
| Latent Heat: $L_{melting}$ | 71 | kJ.kg⁻¹ |
| Melting temperature | 23.4 | °C |
| Descriptive properties | | |
| Thickness | 5.26 | mm |
| Width | 1000 | mm |
| Length | 1198 | mm |

Values of the heat capacity in each phase and melting point have been determined by DSC (Differential Scanning Calorimetry) measurements. These parameters are exposed in details in refs [20,21].

### 4.4. Mathematical model for phase change material

The solidification and melting process are the most studied in building applications. Usually, numerical modelling of these phenomena is either based on the first law of thermodynamics or second law of thermodynamics. For more details, the interested reader may refer to [22].

The thermal model for phase change is based on the apparent heat capacity method from enthalpy method. This method allows to obtain the general form of a heat conduction equation with a nonlinear specific heat, without need to know ahead of time the location of the phase interface. To simplify the mathematical model, some assumptions were made [13,14,23]. Through the solid (or liquid) fraction term, called $f_s$, the final expression of transient heat conduction can be written in 1-D along the $\vec{x}$ direction as follows [13,23]:

$$C_{app}(T)\frac{\partial T(x,t)}{\partial t}=\lambda_{PCM}(T)\frac{\partial^2 T(x,t)}{\partial x^2}, \qquad (2)$$

With:
$$
\begin{cases}
C_{app}(T) = \varrho_s c_s + \Delta(\varrho c) f_s + \dfrac{df_s(T)}{dT}\left(\varrho_l L_m + \Delta(\varrho c).\left(T(x,t) - T_m\right)\right) \\[2mm]
\Delta(\varrho c) = \varrho_l c_l - \varrho_s c_s \\[2mm]
\lambda_{PCM}(T) = \left(1 - f_s(T)\right)\lambda_s + \lambda_l f_s(T) \\[2mm]
f_s(T) = \dfrac{1}{2} - \dfrac{1}{2}\tanh\left(\gamma\,\dfrac{T_m - T(x,t)}{4\delta T}\right) \\[2mm]
\dfrac{df_s(T)}{dT} = \dfrac{\gamma}{8\delta T\left[\cosh\left(\gamma\dfrac{T_m - T(x,t)}{4\delta T}\right)\right]^2}
\end{cases}
$$

The governing equation in terms of the heat apparent capacity can be solved using a standard heat transfer code, and a wide range of discretization approach can be used. As a result, one-dimensional finite difference method can be chosen [19]. To definitely be in accordance with the formalism of ISOLAB equation, the use of a backward Euler scheme is possible thanks to the solid (or liquid) fraction term. Generally, the expression of solid fraction is given by using an approximation of the Heaviside function. In PCM model, a specific parameter called $\gamma$ appears in the final expression. The latter is usually equals to 1, for instance in [24]. In order to not underestimate or overestimate the latent heat value during the phase change process, a proposed method was given by [13]. With this aim, for a given phase change interval $\delta T$, and at $T = T_m$, the expression of the apparent heat capacity ($C_{app}$) leads to the first determination of $\gamma$, according to the following process [13]:

If $\;\boldsymbol{T(x,t) \geq T_m - \delta T}$ **and** $\boldsymbol{T(x,t) \leq T_m + \delta T}$ **then**

$$
\left|\, max(\gamma) = \left(max(C_{DSC}) - \left(\frac{c_s + c_l}{2}\right)\right).\frac{8\delta T}{L_m}
\right.
$$

else

$\left|\,\boldsymbol{\gamma}\;\right.$ must be determined

end

Where $max(C_{DSC})$ is given by DSC, and $\delta T$ is chosen very small. In the numerical simulation, $\delta T = 0.01°C$. In the first approach, $\gamma$ is evaluated as follows:

$$
\gamma = \frac{L_m}{C_{ps}\cdot(T_m - \delta T)} \tag{3}
$$

### 4.5. Main results

The thermal model will be validated only if the validation criteria are reached. These criteria are given by [13,18]:

1. To reach 10% validation error between numerical solutions and experimental data from actual building;
2. To reach acceptable absolute errors between some numerical solutions and measurements:
- ± 2°C for temperatures on each side of each suspended ceiling
- ± 1°C for indoor air temperature of real building

To definitely validate the numerical thermal model, different steps are required. These steps are illustrated in Figure 4:

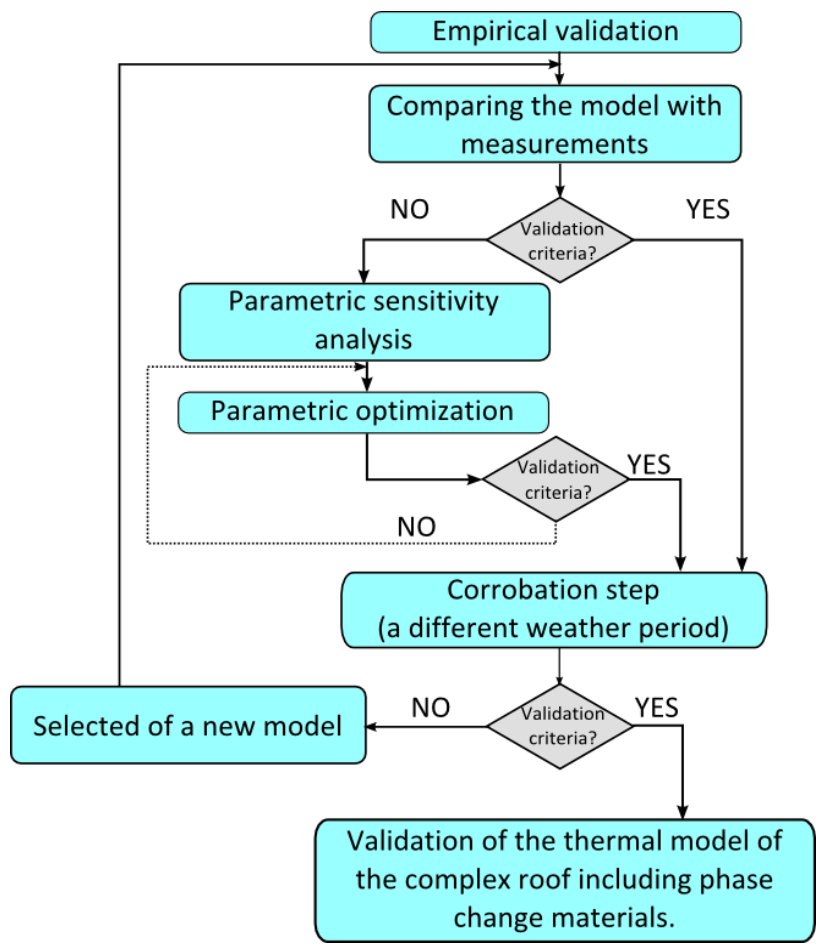

**Figure 4**: Methodology used for the empirical validation

The two first steps have already been presented in details in [13]. The results showed that the numerical thermal model was able to predict the dynamic thermal behavior of PCM. Nevertheless, the validity criteria were not respected. According to the validation methodology, it was necessary to highlight the origins of errors. A parametric sensitivity analysis, according to a method derived from the FAST method (Fast Fourier Amplitude Transform) has been performed and made it possible to put in evidence the parameters with most influence on model outputs [25–27]. Following the sensitivity analysis, the main cause of any difference between the model and experimental data can be explained, but also we can focus on the search of the set of unknown parameters of the model in a restricted range. The following important results were showed [13]:

- The thermal behavior of the complex roof is governed by convective heat transfers, both in the air layer and the faces of the suspended ceiling.
- The absorptivity coefficient of corrugated iron was not the same as new sample.
- The specific parameter $\gamma$ can be considered as a non-dimensional velocity of phase change because the phase change is assumed to occur slowly. This parameter can also be used to overcome numerical instabilities in the zone near the interface between the two phases of the PCM. Indeed, during the phase change, a mushy zone between the two phases is created. From a numerical point of view, sharp discontinuities at the phase interface are observed, implying some numerical instabilities very difficult to overtake. Moreover, it influences the derivative gradient of the solid fraction. However, its physical interpretation is still in investigation.

The parameters linked to the convective exchange coefficients are depicted in Figure 5 and all parameters are given by the Table 3. However, the unknown parameters have to be determined. For this, optimization sequences were performed.

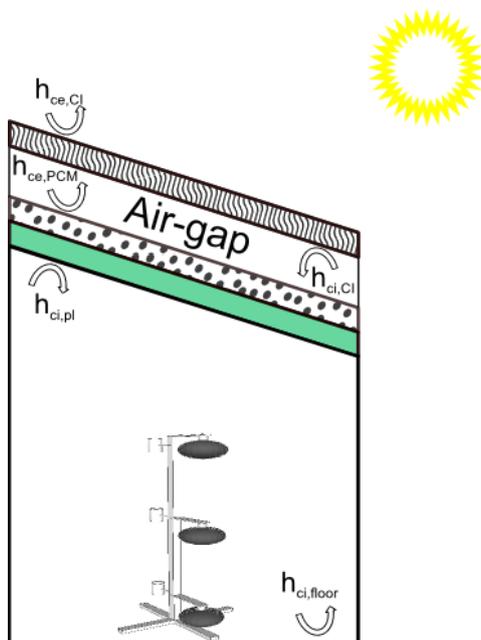

**Figure 5.** Most influential convective exchange coefficients

**Table 3**. Most influential parameters on model outputs at the end of sensitivity analysis step.

| Frequency | Parameters |
|---|---|
| 103 ($h_{ci,floor}$) | Indoor convective exchange coefficient of the floor (zone 1) |
| 107 ($h_{ci,pl}$) | Indoor convective exchange coefficient of the plasterboard |
| 108 (hci,PCM) | Outdoor convective exchange coefficient of the PCM panel |
| 154 | $\gamma$ coefficient |
| 199 ($h_{ci,CI}$) | Indoor convective exchange coefficient of the corrugated iron |
| 200 ($h_{ce,CI}$) | Outdoor convective exchange coefficient of the corrugated iron |
| 218 ($\alpha_{CI}$) | Solar absorptivity coefficient of the corrugated iron |

## 5. Empirical validation of PCM model

### 5.1. PCM model optimization using GenOpt®

The optimization sequence consists in reaching the best set of parameters by maximizing or minimizing a chosen cost function, subject to a set of constraints (or a defined domain), until optimization criteria are reached. Indeed, many simulations are run and when a stopping criterion is reached, like the required error on the studied output for instance, the optimization is stopped.

To optimize the thermal model of PCM implemented in ISOLAB by using GenOpt®, a methodology for the coupling of the building simulation code with the optimization program is required. The study has already been presented and the interested reader may see [18] for details.

Among methods of fixing sets of parameters implemented in GenOpt®, the GPS (Generalized Pattern Search) Hooke-Jeeves MultiStart Algorithm was used. Indeed, it allows to look over some parameters and to avoid that the result values are obtained from a local minimum value. In addition, this algorithm is compatible with all building thermal insulation code and gives reliable results.

The used cost function is calculated by using a mathematical model tool based on the modified standard deviation:

$$s = \sqrt{\frac{1}{n}\sum_{i=1}^{n}(x_i - \bar{x})^2} \qquad (4)$$

If several model outputs are used to optimize the model, the global cost function is the sum of the absolute values of costs functions for each output. The optimization sequence is stopped, if and only if, same minimum values of the cost functions were reached for 10 different optimization sequences. In our approach, several optimization sequences (1500 optimization sequences were reached approximately) and statistics studies were performed.

*5.2. Results of the optimization.*

Results from the optimization sequence are summarized in Table 4.

**Table 4.** Parameters before and after optimization sequences

| Parameters | Before optimization | After optimization |
|---|---|---|
| $h_{ci,floor}$ | 3.50 | 5.00 |
| $h_{ci,w}$ | 1.00 | 1.00 |
| $h_{ci,PCM}$ | 1.00 | 1.50 |
| $\gamma$ (when $T \neq T_m$) | 0.04 | 0.01 |
| $h_{ci,CI}$ | 3.50 | 1.75 |
| $h_{ce,CI}$ | 25 | 5.7V+11.4 [15] |
| $\alpha_{CI}$ | 0.85 | 0.76 |

Each value of an optimized parameter is found in agreement with the physical phenomena. Indeed, convective exchange coefficients are very small and correspond to the non-ventilated upper air layer as the building was kept closed during the experimental sequence. To apply the process for determining $\gamma$ parameter, this coefficient has been evaluated to 0.01 when $T \neq T_m$. The value of absorptivity coefficient ($\alpha_{CI}$) is equal to the value that was determined by [28]. This value, lower than initially expected, can be justified by the   fact that the cell was erected and covered a few years ago, time tending to decrease absorptivity or dark color roof. Indeed, over time, the performances of buildings will be not be the same than when the building was new [15].

*5.3. Comparison between optimized thermal model of PCM with measurements*

On the whole period, the dynamic behaviour of the components of the complex roof and the indoor air temperature are well predicted. Through the different curves (see Figure 6), we can ensure that the given melting temperature for this application was correctly chosen because the temperatures on both sides of PCM allow the storage and the release of energy during phase change process.

The surface temperature of suspended ceiling is predicted at ± 1.3°C (see Figure 7). The used conditions show that PCM model is able to properly predict the behavior of the PCM panel. The comparison between numerical simulations and experimental data shows a good agreement. The indoor air temperature of LGI test cell is predicted with an accuracy of ± 0.5°C (see Figure 8). The criteria for the validation of the thermal model are respected.

According to Table 5, the mean errors between numerical solutions and experimental measurements do not exceed 5%.

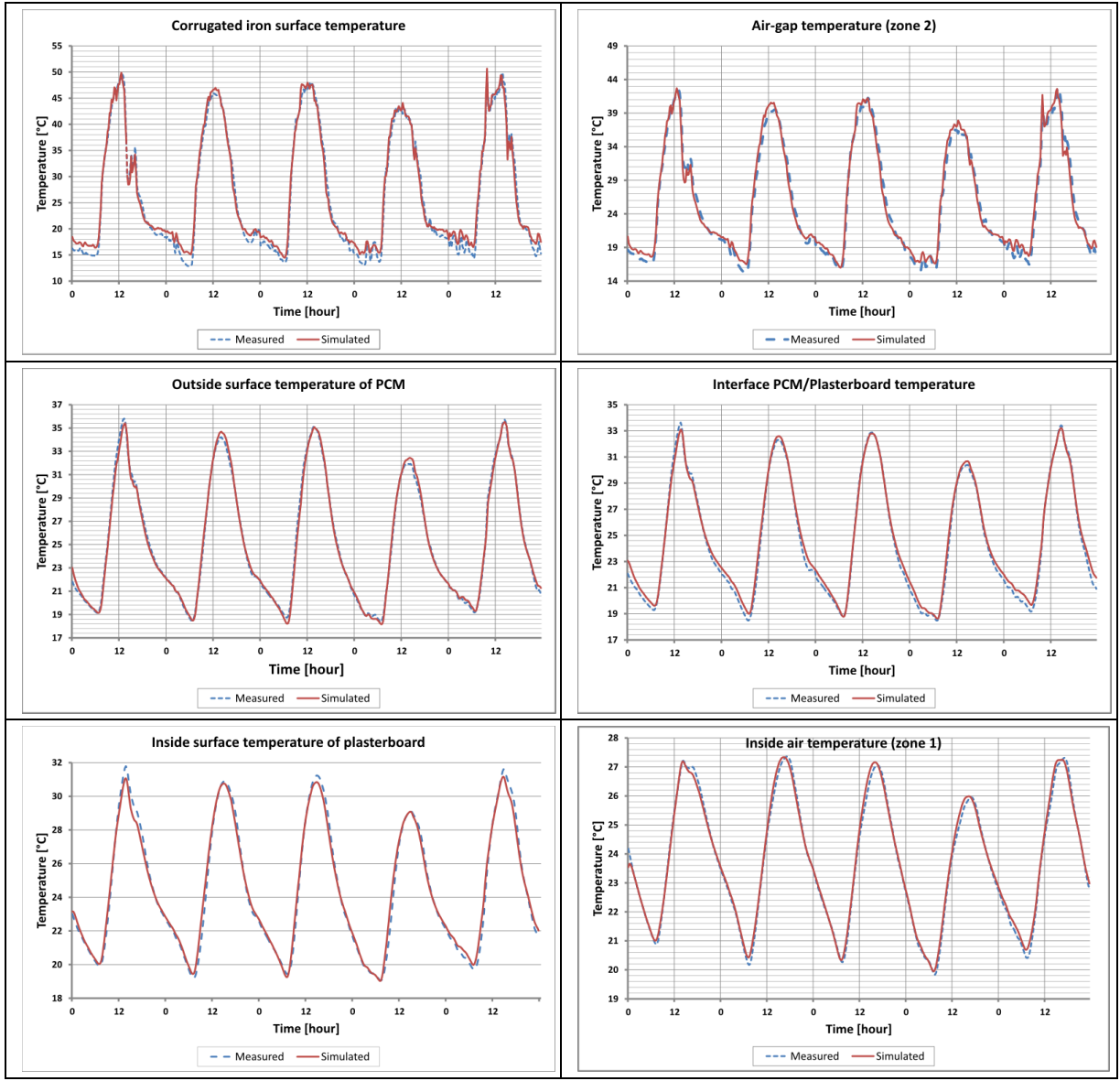

**Figure 6.** Comparison of model results with experimental measurements for all components of the complex roof and for inside air temperature (zone 1)

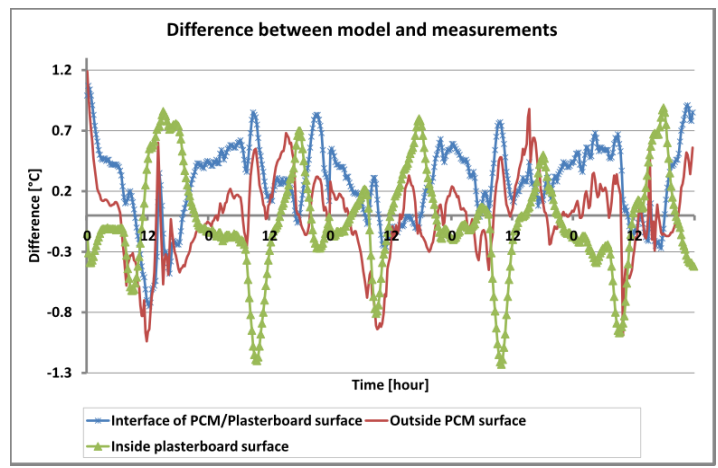

**Figure 7.** Errors between numerical solutions and experimental data for suspended ceiling temperatures.

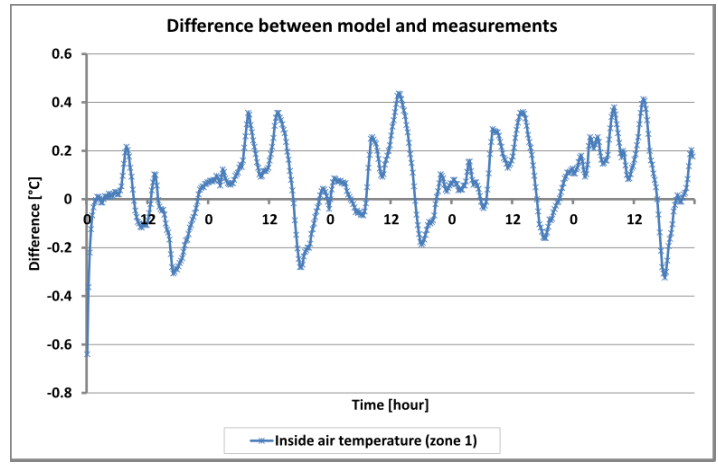

**Figure 8.** Errors between numerical solutions and experimental data for inside air temperature

**Table 5.** Standard deviations, maximum differences and errors after optimization sequences.

| Localization | Standard deviation $\sigma$ [°C] | Maximum difference [°C] | Maximum error [%] | Mean error [%] |
|---|---|---|---|---|
| Inside air temperature (zone 1) | 0.2 | -0.5 | -6.2 | 2.1 |
| Suspended ceiling inside surface temperature (plasterboard) | 0.4 | -1.1 | -8.9 | 2.9 |
| Interface PCM/plasterboard temperature | 0.4 | -1.3 | 8.8 | 3.1 |
| Suspending ceiling outside surface temperature (PCM) | 0.5 | -1.7 | -9.8 | 2.3 |
| Air-gap temperature (zone 2) | 1.0 | -2.6 | -9.4 | 3.1 |
| Corrugated iron temperature | 1.3 | 3.5 | 9.4 | 3.2 |

The most important errors result from the prediction of air-gap temperature and the prediction of the metal sheet temperature in zone 2. For the air-gap, the model must be improved and models from literature based on CFD (Computational Fluid Dynamics) models showed that the results are inconclusive due to the influence of boundary conditions. So, an empirical correlation from experimentation will have to be determined. Nevertheless, the values of convective exchange coefficients ($h_{ce,CI}$ and $h_{ci,PCM}$) determined by the optimization sequences correspond with the empirical correlation developed by Alamdari and Hammond [29] (it has been verified later that the obtained results fit the correlation). The proposal of these authors is a combination of two correlations for taking into account both natural convection in laminar regime and turbulent regime:

$$h = \left[ \left( 1.51 \frac{|\Delta T|^{\frac{1}{4}}}{H} \right)^6 + \left( 1.33 |\Delta T|^{\frac{1}{3}} \right)^6 \right]^{\frac{1}{6}} \quad (5)$$

Where $\Delta T$ is the averaged difference temperature between the surface of the wall and the indoor air of the room. $H$ corresponds to the height of the wall.

With the predicted air temperature in zone 2, the errors between the curves can also be explained by the prediction of the corrugated iron surface temperature. Indeed, the latter has a direct effect on the temperature of each component of the complex roof. To reduce these errors, the radiative model should be improved and the radiosity method should be used [15].

To conclude this step, the thermal model was fully-coupled with ISOLAB code and the results of PCM model are very encouraging. Indeed, for different type of walls with PCM or not, the model is able to predict temperatures in actual conditions. However, to definitely validate the PCM model implemented in the building simulation code ISOLAB, it is necessary to compare the numerical solutions with another experimental data sequence as presented in the next part.

### 5.4. The corroboration step

The corroboration step consists in using another experimental period in order to verify if all parameters determined can be generalized and are not specific for a given period. Moreover, it also allows evaluating the efficiency of the proposed model. For this step of corroboration, the meteorological data used are from October 2 to 5, 2012. This period was characterized by [19]:

- an average maximal global radiation of 900 W.m$^{-2}$
- an average wind speed of 3.50 m.s$^{-1}$
- an average outdoor temperature of 26°C
- an average rate humidity of 65%

The results from the corroboration step are summarized in Table 6. In this part, the validation criteria are also respected. For instance, comparison between numerical simulations and measurements for the predicted indoor air temperature is depicted in Figure 9. For more details, the interested reader may refer to [19].

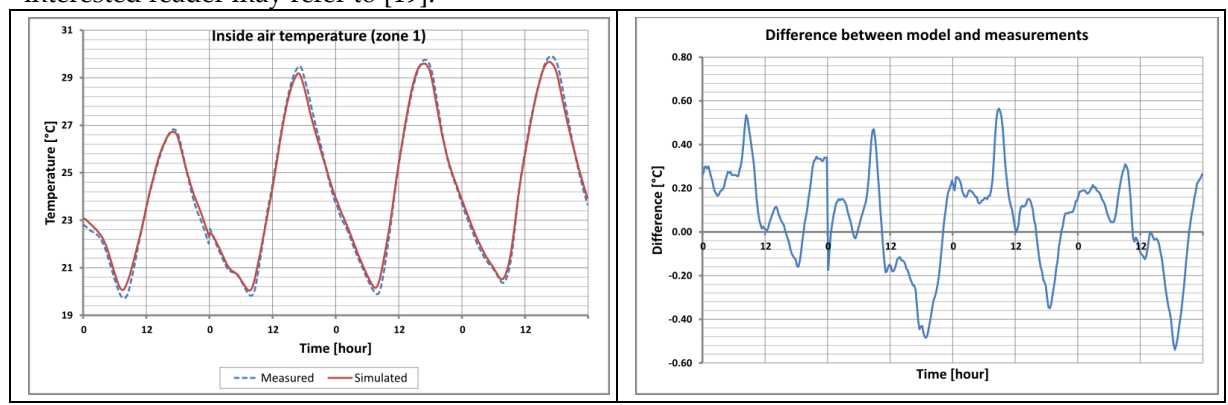

**Figure 9.** Comparison and errors between model results with experimental measurements for inside air temperature (zone 1) during the corroboration step.

**Table 6.** Standard deviations, maximum differences and errors for the corroboration period.

| Localization | Standard deviation $\sigma$ [°C] | Maximum difference [°C] | Maximum error [%] | Mean error [%] |
|---|---|---|---|---|
| Inside air temperature (zone 1) | 0.2 | 0.6 | 5.6 | 2.1 |
| Suspended ceiling inside surface temperature (plasterboard) | 0.5 | 1.5 | 9.2 | 3.4 |
| Interface PCM/plasterboard temperature | 0.5 | 1.7 | 8.9 | 3.6 |
| Suspending ceiling outside surface temperature (PCM) | 0.5 | 1.7 | 9.9 | 2.8 |
| Air-gap temperature (zone 2) | 1.3 | 3.4 | 8.9 | 3.2 |
| Corrugated iron temperature | 1.6 | 4.7 | 9.8 | 3.3 |

Table 5 and Table 6 show that maximum differences between predictions and measurements are a little more important than the results of the first experimental sequence. A possible explanation is a high setting of parameters for specific environmental conditions. Nevertheless, despite these differences the model can be considered as validated because the validation criteria have been reached again. Furthermore, the maximal standard deviation is approximately of 1.6°C and all mean errors are below 5%.

## 6. Conclusions and further works

In this paper, a generic thermal model of PCM in building, validated with reliable experimental data, was presented. An actual building equipped with PCM in the complex roof was set up and each component's surfaces in contact with the indoor air temperature were measured. A detailed investigation was carried out to evaluate the efficiency of the thermal behavior of the model, with important steps included in experimental validation.

A mathematical model based on the apparent heat capacity has been presented. Moreover, an approach of nodal description of the complex wall and a finite difference method in one-dimensional were used.

Thanks to parametric sensitivity analysis, the most influential factors on model outputs such as, all convective exchange coefficients, $\gamma$ parameter and the absorptivity coefficient of corrugated iron, were determined by a generic optimization tool. Then, a new comparison between the optimized thermal model and experimental data was performed. With the corroboration step, the results showed that the validity criteria were respected. Finally, the validation process of model has been reached and the thermal model of PCM has been validated.

Despite the empirical validity of PCM's model, further works are necessary to improve the prediction of air-gap temperature and the radiative model used in order to have obtain better agreement between numerical solutions and measurements. Moreover, a mathematical formulation of $\gamma$ parameter should be presented. Other experimental studies should be led to confirm the use of these materials in tropical climates as Reunion Island. Future works will also focus on the comfort study from the dedicated test cell.

### Acknowledgments

The authors wish to thank Fonds Social Européen and La Région Réunion for their support and their fundings of the first author's thesis. In addition, this research received fundings from the Ministère de l'Outre-Mer.

### Author Contributions

Stéphane GUICHARD designed the experiments and developed PCM model for the experimental and numerical studies. Frédéric MIRANVILLE, Dimitri BIGOT and Harry BOYER helped for the coupling model of PCM with ISOLAB code and performed the validation stages of this one. Bruno MALET-DAMOUR and Teddy LIBELLE helped to monitor the data from the experimentation in order to ensure they were reliable for the empirical validation of the PCM model.

### Conflicts of Interest

The authors declare no conflict of interest.

 <span style="color:red">Draft article submitted and accepted to Energies</span>